\documentstyle[pra,aps,multicol,psfig]{revtex}
\begin{document}
\draft
\title{Vortex-line liquid phases: longitudinal superconductivity
in the lattice London model}
\author{T.J.~Hagenaars$^1$, E.H.~Brandt$^2$, R.E.~Hetzel$^{1,*}$, W.~Hanke$^1$, M.~Leghissa$^{3,**}$, and G.~Saemann-Ischenko$^{3}$\\ 
{\it $^1$Institut f\"ur Theoretische Physik, Universit\"at W\"urzburg\\
Am Hubland, 97074 W\"urzburg, Germany}\\
{\it $^2$Max-Planck-Institut f\"ur Metallforschung, Institut f\"ur Physik\\
Postfach 800665, 70506 Stuttgart, Germany}\\
{\it $^3$Physikalisches Institut III, Universit\"at Erlangen-N\"urnberg\\
Erwin-Rommel-Strasse 1, 91058 Erlangen, Germany}}
\maketitle
\begin{abstract}
We study the vortex-line lattice and liquid phases   
of a clean 
type-II superconductor
by means of Monte Carlo simulations of the lattice London model.
Motivated by a recent controversy regarding 
the presence, within this model, of a vortex-liquid regime with longitudinal
superconducting coherence over long length scales, we directly compare
two different ways to calculate the longitudinal coherence.
For an isotropic superconductor,
we interpret our results in terms of a temperature regime within the
liquid phase in which 
longitudinal superconducting coherence extends over length scales
larger than the system thickness studied. We note that this regime disappears 
in the moderately anisotropic
case due to a proliferation, close to the flux-line lattice melting temperature,
of vortex loops between the layers.
\end{abstract}
\begin{multicols}{2}
\section{Introduction}
The effect of strong thermal fluctuations
on the flux lines is an important aspect of the physics
of high-$T_c$ superconductors in the mixed state.
It is believed that these fluctuations melt the ground-state
flux-line lattice into a vortex-line liquid below
$T_c$ \cite{Blatter,Brandt}.

In Nelson's analogy of vortex lines with world lines of 2D bosons
\cite{Nelson}, the absence of superfluidity of  the bosons
corresponds to the presence of superconducting
coherence along the field direction in the
vortex-line system
\cite{taochen,Feigelman}.
The liquid phase without longitudinal superconductivity
(sometimes referred to as an ``entangled vortex liquid") corresponds
to the 2D boson superfluid state. Similarly, the  vortex-line liquid 
with longitudinal superconductivity (``disentangled vortex liquid")
corresponds to the normal liquid state of the 2D bosons.
The quantity corresponding to the inverse 2D-boson temperature
in the vortex-line system is the system size $L_z$ along the field
direction. Therefore, for a small enough $L_z$, Nelson's analogy predicts
longitudinal superconducting coherence in the vortex liquid. 
In Ref. \cite{Feigelman} it has been argued that the 
disentangled vortex liquid exists as a true
thermodynamic phase $(L_z\rightarrow\infty)$,
at least for the case where the distance
between the vortex lines 
is (much) smaller than the magnetic penetration depth $\lambda_1$.
The scenario in which the longitudinal superconductivity
is destroyed already within the vortex-lattice phase, leading
to a ``supersolid" phase, has been considered in
\cite{Frey}.

There exists a substantial recent literature that deals with
MC simulations of the vortex-line phases, using various model approximations
\cite{taochen,Yinghong,Yinghong2,Xing,Ma,Ryu,Sasik,Sasik2,CaPRL95,CaPRB96,Nguyen}.
In two recent papers, in which the vortex-line system was studied 
within lattice London models \cite{taochen,CaPRL95}, opposite
conclusions have been reached regarding the 
presence of longitudinal superconductivity within
the vortex-liquid phase for systems with $L_z\sim 15-30$ lattice constants.
Although in both works cited, the longitudinal
coherence was argued to exist only over a finite length scale
(and therefore no true disentangled vortex-liquid phase was claimed to exist),
a difference of two orders of magnitude in the results for
this length scale translates into opposite predictions
for the experimentally observable behavior in typical samples
studied in recent flux-transformer experiments, for temperatures
above but close to melting.
One of the motivations of our work has been to clarify the 
reason(s) for the above disagreement.
On the one hand, Chen and Teitel \cite{taochen}, who perform 
the  Monte Carlo simulation at constant magnetic induction $B$, 
found that longitudinal
superconductivity persists far into the vortex liquid for $L_z=30$ lattice
constants.
These authors argue that the length scale beyond which the
vortex lines become entangled,
corresponds to approximately 410 lattice constants.
For YBCO this length scale for longitudinal coherence was estimated
in \cite{taochen}
to be much larger than the thickness of the samples studied in
recent flux-transformer experiments \cite{Safar,Lopez}.
As a result, their findings suggest the existence of a temperature
regime in these experiments, in which the vortex lattice has melted
but in which the flux motion is still maximally $z$-correlated,
in agreement with the results obtained for twinned YBCO samples \cite{Safar}.
On the other hand
Carneiro \cite{CaPRL95} 
found evidence
for the disappearance  of the vortex-liquid regime with
disentangled lines already between $L_z=6$ and $L_z=12$.
These results were argued to agree with later experiments on untwinned YBCO
samples, in which the flux-line lattice melting and a loss
of maximally $z$-correlated vortex motion was found to coincide.

In Carneiro's calculations the longitudinal response
is calculated in a different way,
discussed in more detail below,
in which fluctuations in the net vorticities (in two directions) are allowed.
In a recent Comment \cite{PRLComment}, Chen and Teitel
have criticized Carneiro's calculational scheme. They suggested
that it effectively measures the transverse instead of the 
longitudinal response.
Below, we will
present results of MC simulations of the lattice London model.
We present a direct comparison of the  two different calculational schemes,
and will point out that the method of Ref.~\cite{CaPRL95} has some serious
problems.

From our results we conclude that the
isotropic lattice London model predicts a
temperature regime in
which longitudinal coherence over long length scales exists
within the vortex-liquid phase.
We also present results for
a moderately anisotropic system, in  which the temperature
at which longitudinal coherence is lost,
is roughly equal to the melting temperature.

The outline of this paper is as follows.
In section II we introduce the model and the Monte Carlo
method(s).
In section III we present our results.
We summarize our findings in section IV. 

\section{Lattice London Model and MC Algorithms}

The first MC study of the lattice London model was performed by
Carneiro, Cavalcanti and Gartner \cite{CaCaval}.
The model was explored further by Carneiro
for a superconductor with free surfaces \cite{CaPRB94}.
The lattice London model is formulated directly in terms
of the vortex degrees of freedom. The vortex lines are
modeled as consisting of elements of unit length
based at dual lattice sites of a cubic lattice.
At every dual lattice site we define three
integers $q_\mu(\bbox{R}) =0, \pm 1,\pm 2,...$,
the vorticities in the directions $\mu =x,y,z$.  
When a vorticity $q_\mu(\bbox{R}_i)$ is nonzero, its magnitude gives the
number of flux quanta carried by the associated 
vortex-line element.
The $q_\mu (\bbox{R}_i)$ are subject to the continuity constraint
\begin{equation}
\sum_{\bbox{e}_\mu}\left[ q_\mu (\bbox{R}_i)-q_\mu (\bbox{R}_i-\bbox{e}_\mu)\right] =0
\label{constraint}
\end{equation}
that ensures that the vortex-line elements form either closed loops
or lines that end at the boundaries. Here $\bbox{R}_i-\bbox{e}_\mu$ runs over
nearest neighbor sites of $\bbox{R}_i$.
At constant $B$, the Hamiltonian
of the isotropic lattice London model, expressed in terms  of
the vorticities, is given by:
\begin{equation}
{\cal H}=4\pi^2 J\sum_{i,j,\mu}q_\mu(\bbox{R}_i)q_\mu(\bbox{R}_j)g_\mu(\bbox{R}_i-\bbox{R}_j)
\label{Ham}
\end{equation}
where $J=\Phi_0^2 d/(32\pi^3\lambda_1^2)$ and $g_\mu(\bbox{R})$ 
is the  London  interaction with Fourier components
\begin{eqnarray}
& &g_{x,y}(\bbox{k})=\frac{R}{\kappa_1^2+\kappa_2^2 +R(\kappa_3^2+ (d/\lambda_1)^2)} \nonumber\\
g_{z}(\bbox{k})&=&\frac{\kappa^2+R(d/\lambda_1)^2}{(\kappa^2+(d/\lambda_1)^2)(
\kappa_1^2 +\kappa_2^2 +R(\kappa_3^2 + (d/\lambda_1)^2))} 
\label{Fcom}
\end{eqnarray}
where $\kappa_\mu^2 =2-2\cos k_\mu$ ($k_\mu =2\pi n_\mu/L_\mu,$ 
$ n_\mu= 0,1,...,L_\mu -1$) and $\kappa^2=\sum_\mu \kappa_\mu^2$.
Here we assumed an $L_x\times L_y \times L_z$ lattice with  
periodic boundary conditions.
$R=\lambda_1^2/\lambda_3^2=1/\Gamma^2$ is the anisotropy parameter, with 
$\lambda_1$ and $\lambda_3$ the magnetic penetration depths along the $x$-$y$ plane
and the $z$ axis, respectively, and $d$ the lattice constant.
The Hamiltonian (\ref{Ham}) can be either derived
from the discrete version of the London free energy \cite{CaCaval,CaPRB94}
or by a duality transformation of the Hamiltonian of
a Villain-type lattice superconductor
\cite{DasHal}.

Monte Carlo sampling of the phase space for the variables
$q_\mu(\bbox{R})$  at constant  $B$
is performed as follows.
The initial configuration is prepared to contain the number of vortex
lines we want to study, depending on the value chosen for
the magnetic induction $B$.
A Monte Carlo 
update step consists of adding at a given site a closed
$d\times d$ square loop of unit vorticity with an orientation chosen
randomly from the six possible ones. The addition of a closed
loop preserves the constraint (\ref{constraint}).
The standard Metropolis algorithm is employed to accept or reject
the new configuration. 
Obviously, when only closed loops are added, the magnetic induction 
$B$ with components
\begin{equation}
B_\mu= \frac{\Phi_0}{d^2 V}\sum_j \langle q_\mu (\bbox{R}_j)\rangle .
\label{Bdef}
\end{equation}
($V=L_xL_yL_z$) will be constant throughout the simulation. 

To probe  superconducting coherence one can consider the helicity moduli
introduced by Chen and Teitel \cite{chenteitel1,taochen}
$\Upsilon_\mu=\lim_{k_\nu\rightarrow 0}\Upsilon_\mu(k_\nu)$. Here $\Upsilon_\mu(k_\nu)$
is the linear response coefficient between
a perturbation $\delta A_\mu^{ext}(k_\nu)\bbox{e}_\mu$ of the external vector potential and
the induced supercurrent
\begin{equation}
j_\mu(k_\nu)=-\Upsilon_\mu(k_\nu)\delta A_\mu^{ext}(k_\nu),
\end{equation}
where $\bbox{k}\cdot \bbox{A}^{ext}=0$ (London gauge).
For our lattice London model these moduli are given by
\begin{eqnarray}
\Upsilon_\mu(k_\nu)=\big.\frac{J\lambda_1^2\kappa^2}{1+\lambda_1^2\kappa^2}
\big|_{k_\mu=k_\sigma=0}\times\gamma_\mu(k_\nu),\nonumber\\
\gamma_\mu(k_\nu)\equiv 1-\frac{4\pi^2J\lambda_1^2}{Vk_BT}\frac{\langle q_\sigma(k_\nu)q_\sigma(-k_\nu)\rangle}{1+(1+\delta_{\mu,z}(1/R-1))\lambda_1^2\kappa_\nu^2},\label{gamma}
\end{eqnarray}
where $(\mu,\nu,\sigma)$ is a cyclic permutation of $(x,y,z)$, $q_\sigma(k_\nu)\equiv q_\sigma(k_\nu,k_\mu=k_\sigma=0)$, and $\langle ..\rangle$ denotes a
thermal average.
We will focus on the quantity $\gamma_z\equiv 
\lim_{k_x\rightarrow 0} \gamma_z(k_x)$. 
We note that for a uniaxial superconductor with $B$
parallel to the $z$ axis, one has $\gamma_z =
\lim_{k_y\rightarrow 0} \gamma_z(k_y)$ 
from symmetry, where $\gamma_z(k_y)$ is defined by
the expression (\ref{gamma}) with the indices $x$ and $y$ interchanged.
If $\gamma_z=1$, one has a perfect screening of the corresponding perturbation
of the vector potential, and thus longitudinal superconducting coherence.
A sharp jump in $\gamma_z$ from unity is interpreted as loss of
longitudinal coherence.
In Ref. \cite{chenteitel1} it has been shown that $\gamma_z$ will be non-zero
as long as the vortex-line system retains a finite shear modulus
at finite wave vector. Physically, this may be understood by
the finding of Ref. \cite{BrandtJLTP} that helical
instabilities of the vortex lines driven by
a longitudinal current, can be stabilized either by pinning
or by a finite shear modulus.

In the MC calculations, $\gamma_\mu$ is estimated from the
quantity $\gamma_z(k_\nu)$ at small $k_\nu$.
In Ref. \cite{taochen} this was done by extrapolation of
the results for $\langle q_\sigma(k_\nu)q_\sigma(-k_\nu)\rangle$
to $k_\nu=0$ using a fit to a polynomial in $\kappa_\nu^2$. 
It turns out that such a fit does not work for higher
values of $\lambda$ and/or $\Gamma$ than considered in \cite{taochen}.
In Ref. \cite{CaPRB96} this has been demonstrated for
$\lambda=12, \Gamma=1$.
However, this problem disappears if one examines the full
ratio that defines $1-\gamma_\mu(k_\nu)$ (see Eq.~(\ref{gamma})). 
This quantity has only a weak dependence
on $k_\nu$ \cite{opmerking0}, so that one already obtains  a reliable estimate
for $1-\gamma_\nu$ from the data point for the smallest nonzero
wave vector, $k_\nu=2\pi/L_\nu$.
This direct way of estimating the response was employed before
in Ref.~\cite{Nguyen}.
We illustrate it with an example in 
Figs.~10 and 11.
In particular, $1-\gamma_z(k_x)$ is almost independent of $k_x$
(when $B$ is along $z$).
Therefore, it is possible to obtain meaningful results
for $\gamma_z$ already for systems with $L_x$ equal to only a few
times the average distance between the vortex lines.

In Ref. \cite{CaPRL95}, a different set up was introduced
for measuring longitudinal superconducting coherence.
This original set up
starts from the observation that {\em at a free boundary}
parallel to the $y$-$z$ plane,
superconducting coherence in the $z$ direction corresponds
to Meissner shielding of an infinitesimal uniform applied field
in the $y$ direction \cite{BrandtJLTP}.
For a lattice London model in the form of a slab, with fbc
in the $x$ direction and pbc in the other directions,
and in a constant {\em applied} magnetic field,
this shielding is measured  by the 
transverse magnetic permeability $\mu_y$.
\begin{equation}
\mu_y= \frac{\partial B_y}{\partial H_y}\big|_{H_y=0} = \frac{Vd^3}{4\pi k_BT}\langle b_y^2\rangle,
\label{lsup}
\end{equation}
where $\langle ..\rangle$ is now a thermal average weighted by the
Hamiltonian
\begin{equation}
\tilde{\cal H}= {\cal H} - \frac{Vd^3}{4\pi}\Phi_0 /d^2  
\sum_{\mu}f_\mu b_\mu .
\label{Ham2}
\end{equation}
Here  $f_\mu =H_\mu d^2/\Phi_0 $
is the dimensionless 
applied magnetic field in units of the elementary
flux quantum $\Phi_0$ per plaquette area.
In (\ref{Ham2}) $f_\mu$ multiplies the microscopic magnetic
field $b_\mu$:
\begin{equation}
b_\mu= \frac{\Phi_0}{d^2V}\sum_{i} q_\mu(\bbox{R}_i).
\label{mmf}
\end{equation}
When $\mu_y$ is zero, the external field $H_y$ is shielded by supercurrents
running parallel to $\bbox{e}_z$ and thus the system is a longitudinal
superconductor. When $\mu_y$ is nonzero, this is interpreted as a
loss of longitudinal superconductivity.
Note that $\mu_y=1-\gamma_z(k_x=0)$, which is identically zero in
the case of periodic boundaries and constant $B$.
For a slab placed in a constant applied magnetic field however,
the vorticity fluctuations measured by the 
thermal average in the right hand side of Eq.~(\ref{lsup}) can be
nonzero by virtue of the free boundaries.
At such boundaries
the elementary vorticity fluctuations  satisfying the
constraint (\ref{constraint})  are 
elementary loops with one
side missing (the side outside the sample), so-called {\em incomplete 
loops}. 

The MC method at constant applied field,
introduced by Carneiro in Ref. \cite{CaPRL95},
uses the vortex-vortex interaction $g(\bbox{R})$ for a system with periodic
boundary conditions in all directions, but 
updates at sites in the planes  $x=1$ and 
$x=L_x$ as if these planes were (adjacent to) free boundaries,
allowing the addition
of closed elementary loops parallel 
to the  $y$-$z$ plane 
and of incomplete loops parallel to the
$x$-$y$ and $x$-$z$ planes.
The periodic boundary conditions are imposed also in the $x$ direction
(through the form of $g(\bbox{R})$)
to diminish finite size effects.
Although $x=1$ and $x=L_x$ are therefore no genuine free
boundaries, Carneiro has claimed that 
the addition of incomplete loops incorporates the essence
needed for the calculation of $\mu_y$.
It is however not a priori  evident whether or not this treatment
of the boundaries leads to meaningful results for $\mu_y$.
In fact, we think that it does not, as we will argue from
a consistency check presented in paragraph III.A.

In Carneiro's method, the allowed incomplete loops
do not lead to net vorticity fluctuations
in the $x$ direction.
We note that this allows us to simultaneously extract 
$\gamma_z$ from a simulation in which we
calculate $\mu_y$, namely 
from the fluctuations of
$q_x$ at $k_x=k_z=0$ and the smallest nonzero $k_y$. 
As a result, we are able to compare the results
of the two methods in one and the same simulation.

\section{Results }
We have considered vortex-line systems
corresponding to $B=1/8$, $B=1/15$ and  $B=1/24$
($B$ along $z$ and measured in units of $\Phi_0/d^2$).
The corresponding ground-state
vortex lattices are shown in Fig.~1.
\begin{figure}
\centerline{
\hbox{ (a) \hspace{2.2cm} (b) \hspace{2.2cm} (c)}
}
\centerline{
\hspace{2.2cm}
\psfig{figure=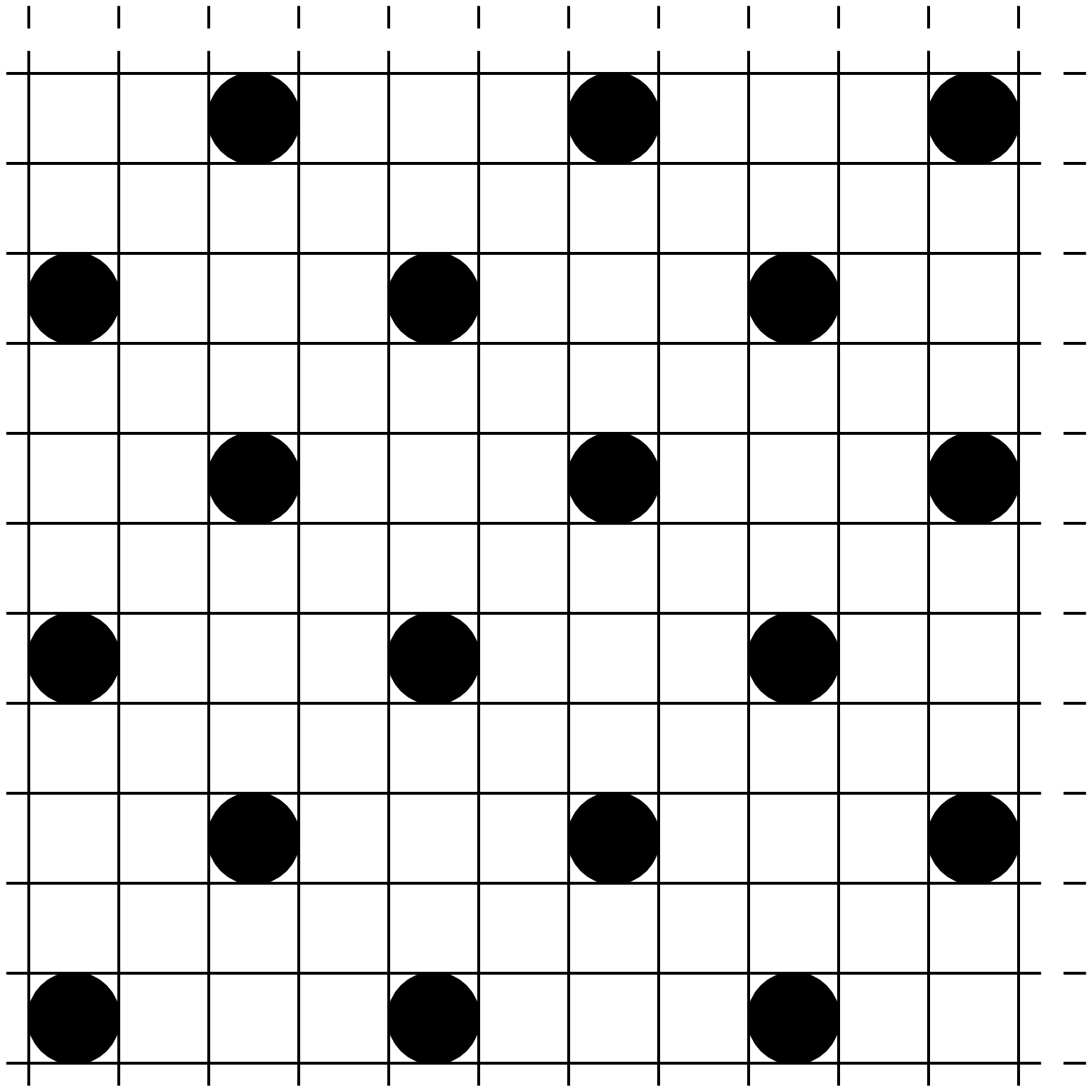,height=2.3cm,width=2.3cm}
\hspace{0.3cm}
\psfig{figure=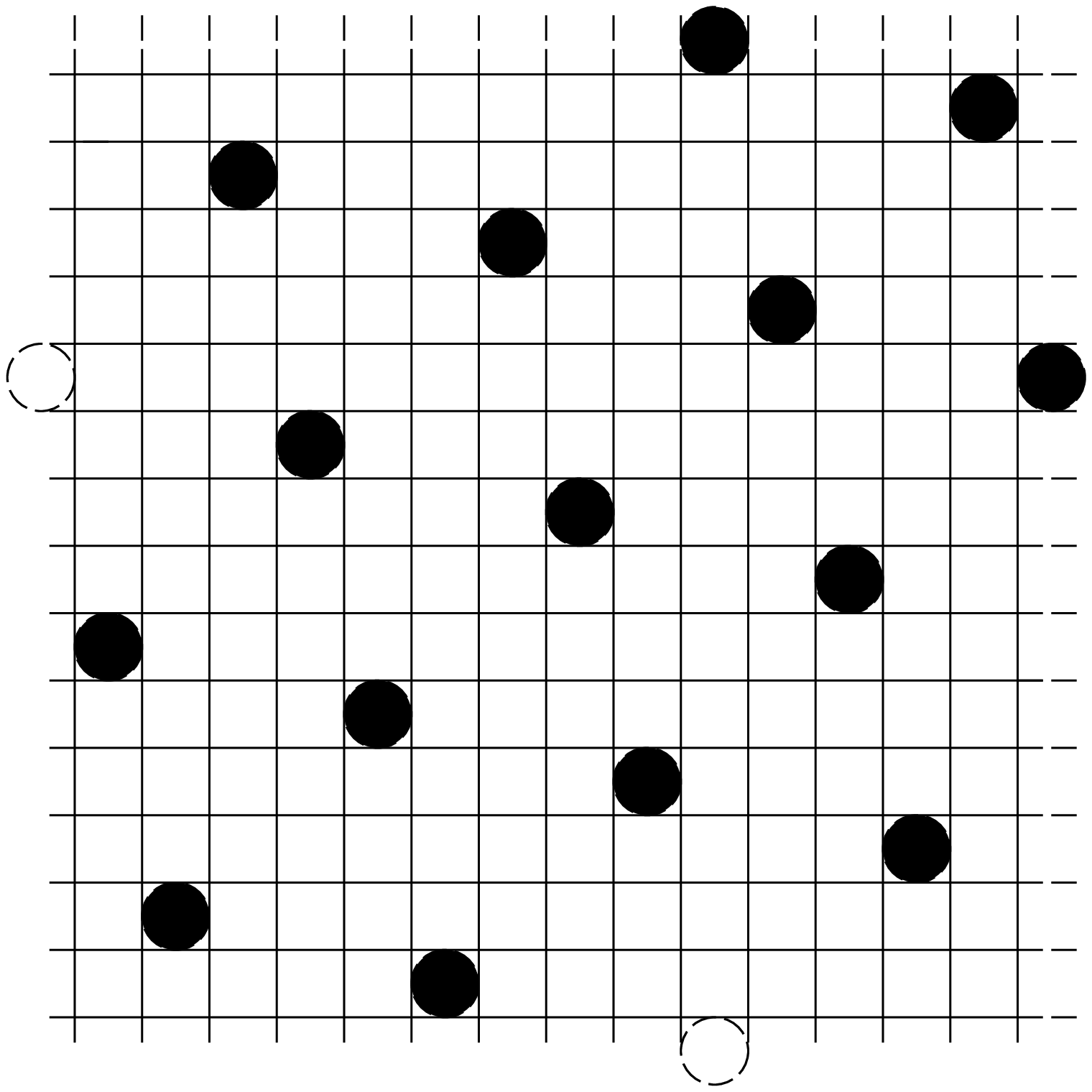,height=2.3cm,width=2.3cm}
\hspace{0.3cm}
\psfig{figure=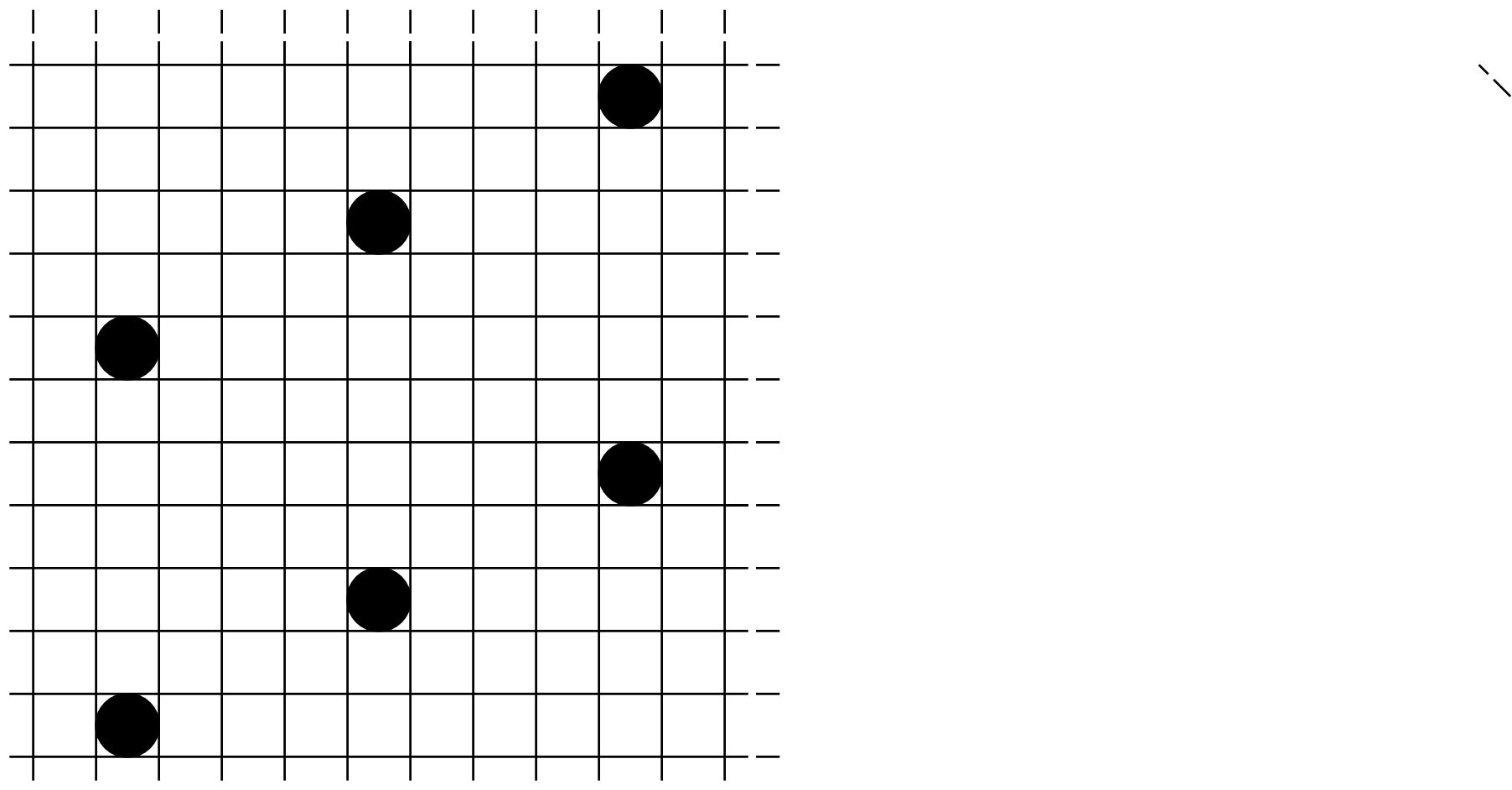,height=2.3cm,width=4.5cm}
}
{FIG.~1. \small Ground state vortex lattices, projected onto the $x$-$y$ plane, for 
different values of the dimensionless magnetic field $f$: (a) $B=1/8$, (b) $B=1/15$, (c) $B=1/24$.}
\label{fig1}
\end{figure}
For each temperature we start in the ground-state
configuration. Runs consist of 16384 MC sweeps through the lattice.
Of these the first 8192 sweeps are discarded for equilibration, and the
second 8192 are used to compute the thermal averages.
\subsection{Isotropic case.}
We first present results obtained employing Carneiro's method, 
for different values of the magnetic field
$f_x=f_y=0, f_z=f$.
For $B=1/8$ and $B=1/24$ we chose the commensurate
lattice size $L_x=L_y=L_z=12$ and for $B=1/15$ the sizes
$L_x=L_y=15, L_z=6,15$. 
In Fig.~2 we show the vortex-lattice correlations
for vortex elements $q_z(\bbox{R}_i)$ in the $z$-direction:
\begin{equation}
P(\bbox{R})=\frac{\sum_j
\langle q_z(\bbox{R}_j+\bbox{R})q_z(\bbox{R}_j)\rangle}{\sum_j\langle
q_z^2(\bbox{R}_j)\rangle}
\end{equation}
as a function of temperature, for
$B=1/8, \lambda_1=12$;
they are in quantitative agreement
with the corresponding results
reported in Refs. \cite{CaPRL95} and \cite{CaPRB96}.
We measure the temperature in units of the coupling $J/k_B$.
The melting of the flux-line lattice is reflected
in the decay of the corresponding correlations.
From the curves in Fig.~2 we may estimate the melting
temperature to be
$T_m=2.7\pm 0.1$. 
This estimate is actually
an upper bound of the true melting temperature.
As has been noted in Refs.~\cite{taochen,Nguyen}, the discrete
mesh introduces an artificial pinning potential for the vortex lines
that tends to increase the apparent melting temperature.
This effect is reduced when considering smaller vortex densities,
that correspond to a finer effective mesh. In \cite{Nguyen} 
the depinning from the mesh was argued to occur
below the (true) melting temperature for $B$ lower than 
approximately $B_c=1/32$.
\begin{figure}
\centerline{
\hbox{
\psfig{figure=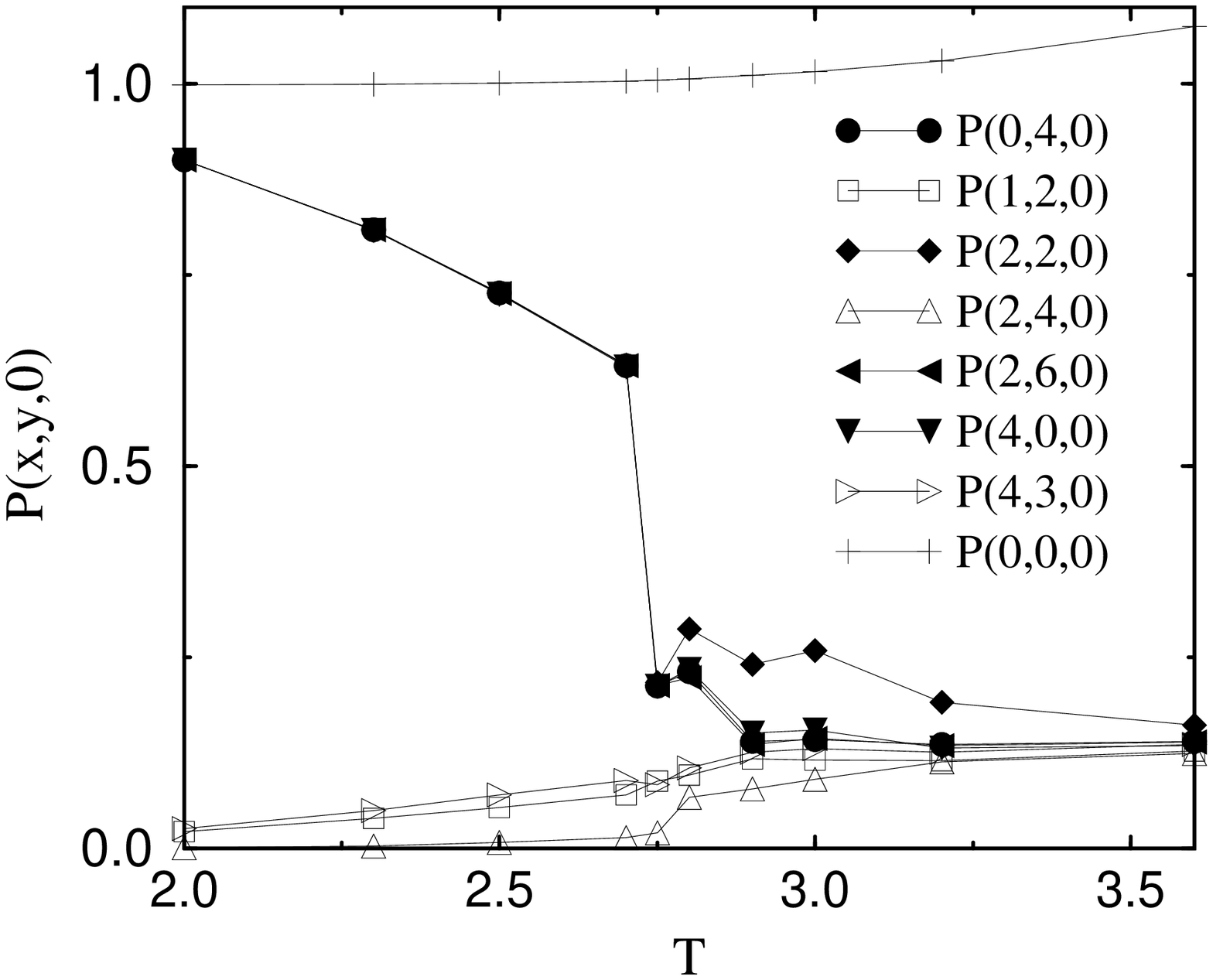,height=6.50cm,width=7.50cm}
}}
{FIG.~2. \small Vortex-vortex spatial correlations $P(\bbox{R})$
for $B=1/8,$ $\lambda_1=12,$ $L_x=L_y=L_z=12$ as a function of temperature.
Temperature is measured in units of $J/k_B$.
The filled symbols represent correlation functions that
measure the translational order corresponding to the ground-state
configuration depicted in Fig.~1(a).}
\label{fig2}
\end{figure}
As an illustration of the liquid phase,
in Fig.~3 a snapshot of the
vortex configuration for 
the $B=1/15$, $\lambda_1=15$, and
$T=2.4=1.2T_m$ is shown.
In this snapshot we observe
a few (vortex-line segments built up of) 
incomplete elementary loops that
were generated in the planes $x=1$ and $x=L_x$.

\begin{figure}
\centerline{
\psfig{figure=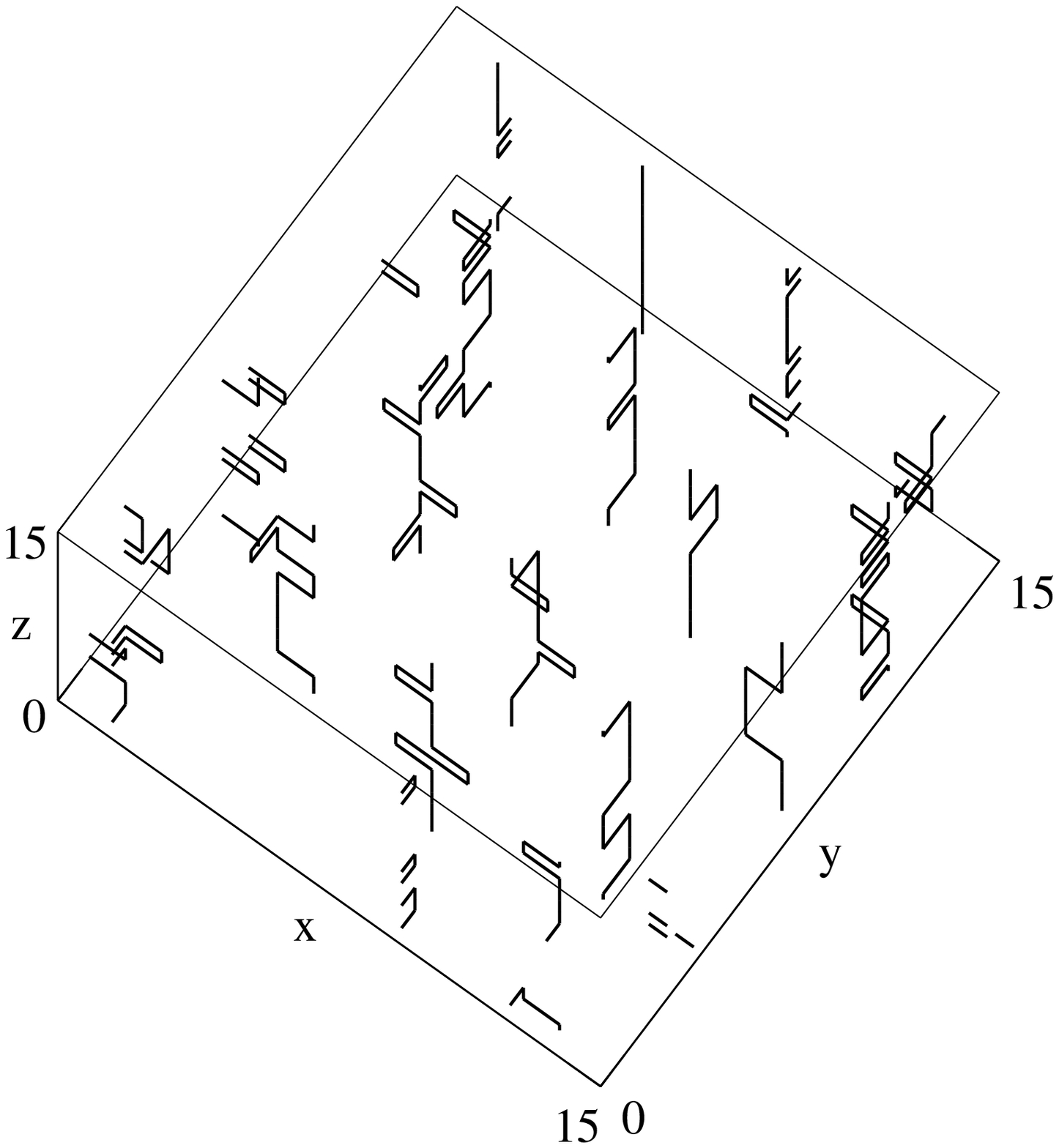,height=5.00cm,width=6.00cm}
}
{FIG.~3. \small Snapshot of the vortex-line configuration in the liquid phase
for $B=1/15,$ $\lambda_1=15,$ $L_x=L_y=L_z=15,$ $T=2.4$. The perspective is from above,
the upper and lower $x$-$y$ planes are enclosed by thin lines,
and the fat lines indicate the vortex lines. Note that periodic
boundary conditions are imposed in all directions, and that at some places
different parts of one and the same vortex line connect through these
boundaries. Due to the introduction of incomplete elementary loop fluctuations
in the planes $x=1$and $x=15$, the microscopic field is allowed to
fluctuate (see text).}
\label{fig3}
\end{figure}
\begin{figure}
\psfig{figure=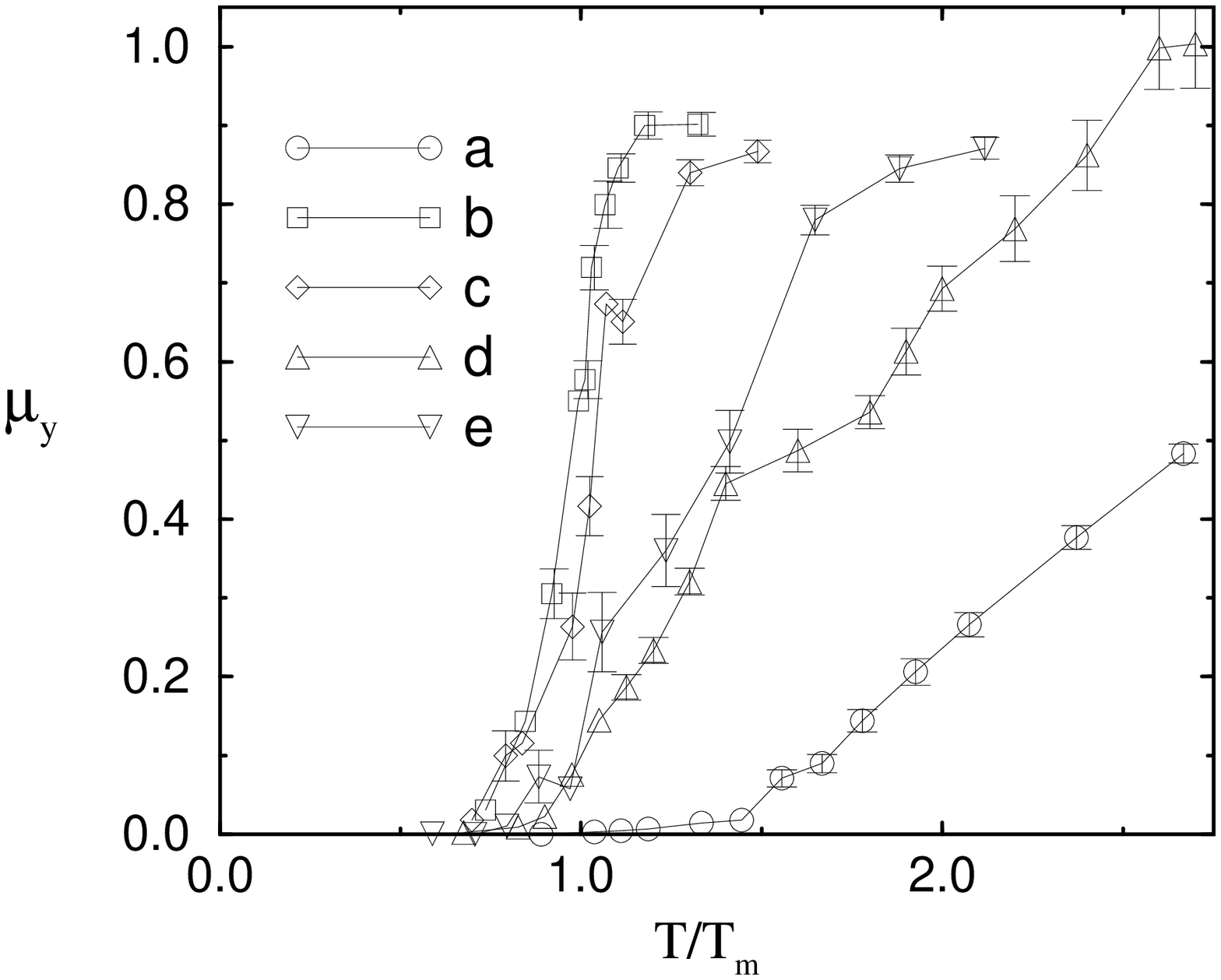,height=7.00cm,width=8.20cm}
{FIG.~4. \small The transverse 
response $\mu_y$ as a function of temperature,
for the parameters (a) $B=1/15,$ $\lambda_1=5,$ $L_x=L_y=15,$ $L_z=6$, (b) $B=1/8,$ $\lambda_1=12,$ $L_x=L_y=L_z=12$, (c)
$B=1/15,$ $\lambda_1=15,$ $L=15$, (d) $B=1/15,$ $\lambda_1=5,$ $L_x=L_y=L_z=15$, and (e)
$B=1/24,$ $\lambda_1=12,$ $L=12$. Sample error bars are shown.
For all curves the temperature is normalized to the 
melting temperature $T_m$.
These melting temperatures are 
$T=1.35\pm 0.1$ (a) and $T=2.7\pm 0.1$ (b),
$T=2.15\pm 0.1$ (c), $T=2.0\pm 0.1$ (d), $T=1.7\pm 0.1$ (e),
and have been estimated from the decay of the
vortex-lattice correlations as shown in Figs. 2 and 3 [25].}
\label{fig4}
\end{figure}

In Fig.~4, curve (b) shows $\mu_y$ as a function of temperature
for $B=1/8, \lambda_1=12$. As in Refs. \cite{CaPRL95,CaPRB96},
$\mu_y$ is non-zero above the melting temperature.
Curve (d) corresponds to $B=1/15, \lambda=5$,
the case studied at constant $B$
in Ref.~\cite{taochen}. 
Here  $\mu_y$ is still small at the melting temperature,
but rises fast for slightly higher temperatures. 
For $L_z=6$ (curve (a)) we find that $\mu_y$ stays close to zero in
a substantial temperature
regime above $T_m$. As has been noted before in
Ref.~\cite{CaPRB96}, this shows that $\mu_y$ does not correspond 
to the transverse response $1-\gamma_x$, as was suggested in 
Ref. \cite{PRLComment}, because the latter response is 
resistive in the liquid phase. 
For further comparison, we have included results
for $B=1/15,\lambda=15$ (c) and $B=1/24, \lambda=12$ (e).
For all cases (b)-(e), with $L_z=12$ or $L_z=15$, $\mu_y$ starts to rise 
at around the melting temperature. 
This is in sharp contrast with the results reported for
$\gamma_z$ in Refs. \cite{taochen,Nguyen}.

We have also extracted $\gamma_z$ from  these same simulations,
in the way explained in Sec II.
In Fig.~5 we compare $\mu_y$ (triangles) and
$1-\gamma_z$ (circles) for
$B=1/15, \lambda_1=5, L_z=15$, the same parameters that were
considered in Ref.~\cite{taochen}.
Up to $T=2.5T_m$, these results for $\gamma_z$ are within error bars
identical to those for $\gamma_z$ calculated using the
algorithm of Ref. \cite{taochen} (shown as squares).
We find that $1-\gamma_z$ rises from close to zero to close to one
at a temperature  approximately twice as large as $T_m$.
Similarly, for all other curves in Fig.~4
with exception of curve (b) ($f=1/8$), we find that
$\gamma_z$ is close to one up to the highest temperature values
shown.
Therefore, the temperature at which $\gamma_z$ drops from unity
lies well above the melting temperature, in agreement
with the results obtained
in Refs. \cite{taochen} and \cite{Nguyen}.
For $B=1/8$ (curve (b) in Fig.~4, 
$\lambda=12$) the temperature
region between the drop of $P(\bbox{R})$ and the drop of $\gamma_z$
is smaller. This is probably caused by 
the pinning artefact due to the discrete mesh,
that tends to increase the apparent melting temperature
more for $B=1/8$ than for the
lower vortex densities studied here\cite{opmerking}.
\begin{figure}
\centerline{
\hbox{
\psfig{figure=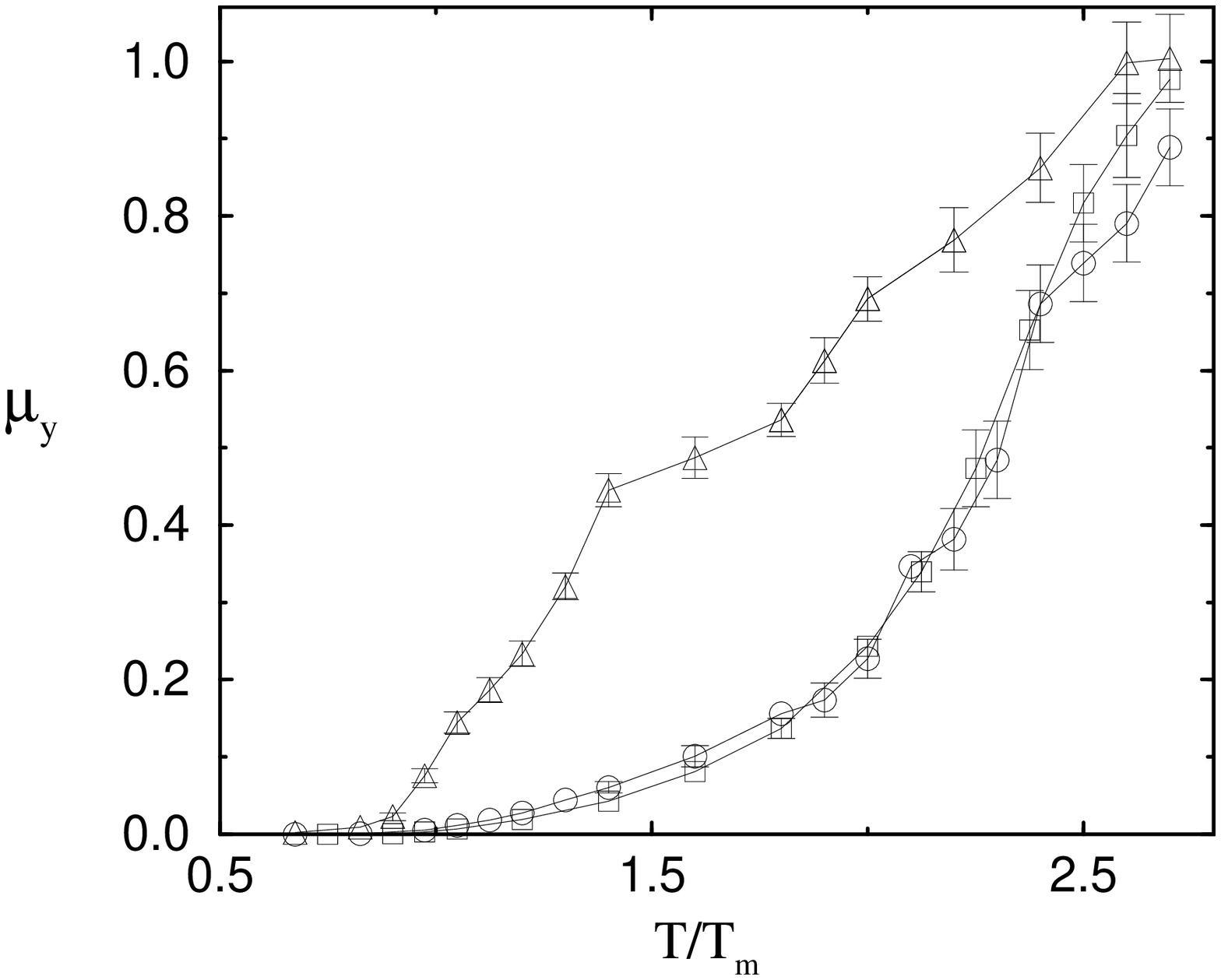,height=6.20cm,width=7.50cm}
}}
{FIG.~5. \small Comparison of the responses $\mu_y$ (triangles) and $1-\gamma_z$
(circles)
for $B=1/15,$ $\lambda_1=5,$ $L_x=L_y=L_z=15$ as a function of temperature.
Temperature is measured in units of $J/k_B$.
The data points indicated with squares are results for $1-\gamma_z$
from an independent simulation using the MC method with
constant $B$. Sample error bars for
$1-\gamma_z$ refer to the error in the data point for the smallest
nonzero $k_\nu$ (see text).}
\label{fig5}
\end{figure}
\end{multicols}
\begin{figure}
\vspace{-0.5cm}
\centerline{
\psfig{figure=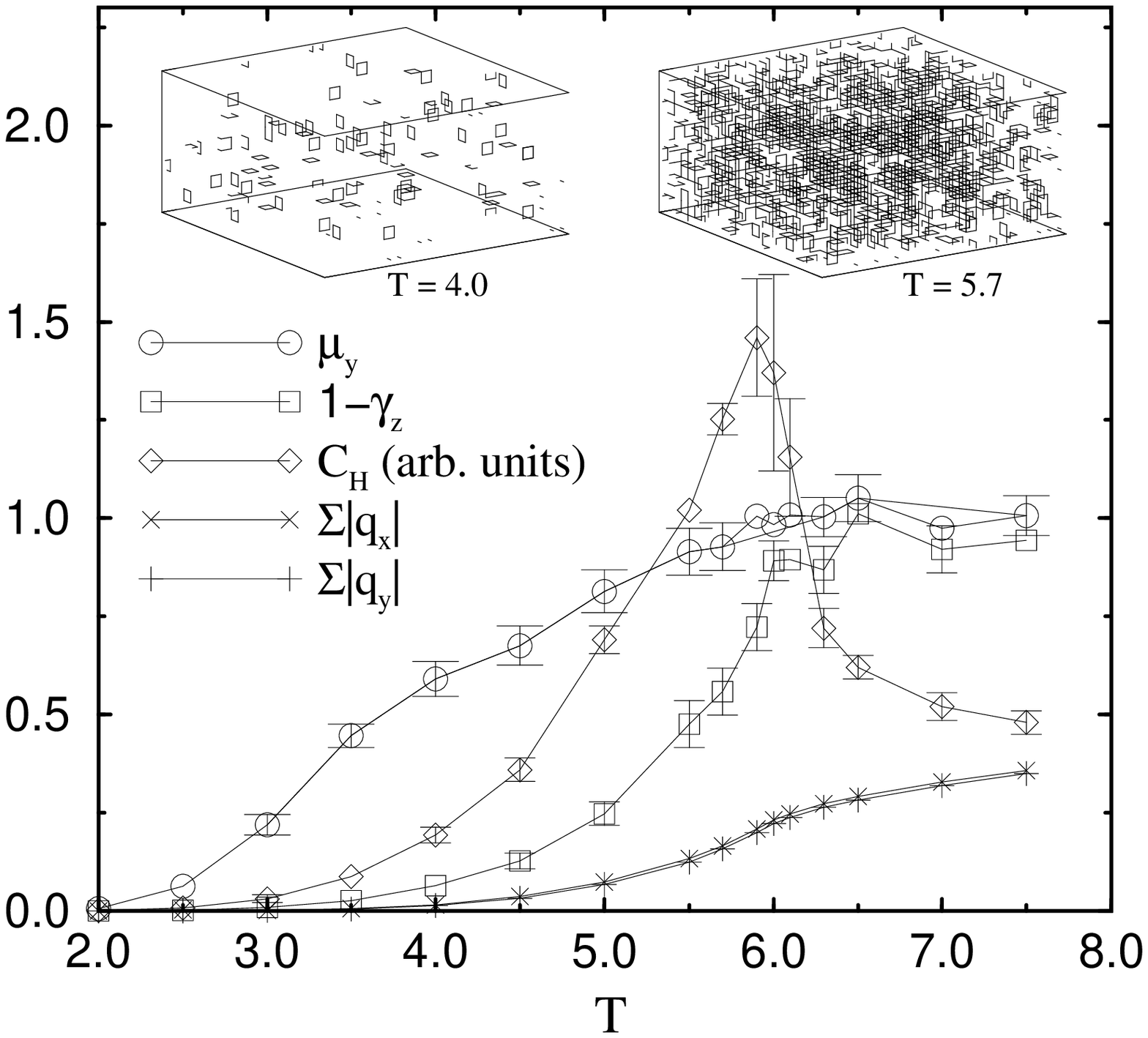,height=9.00cm,width=10.70cm}
}
{FIG.~6. \small Responses $\mu_y,1-\gamma_z$, specific heat $C_H$, and
absolute vorticities
for $B=0,$ $\lambda_1=12,$ $L_x=L_y=24,$ $L_z=12$.
Sample error bars are shown for $\mu_y,1-\gamma_z$, and $C_H$.
The insets show snapshots of the vorticity configurations
at $T=4.0$ and $T=5.7$.}
\label{fig6}
\end{figure}
\begin{multicols}{2}
\begin{figure}
\psfig{figure=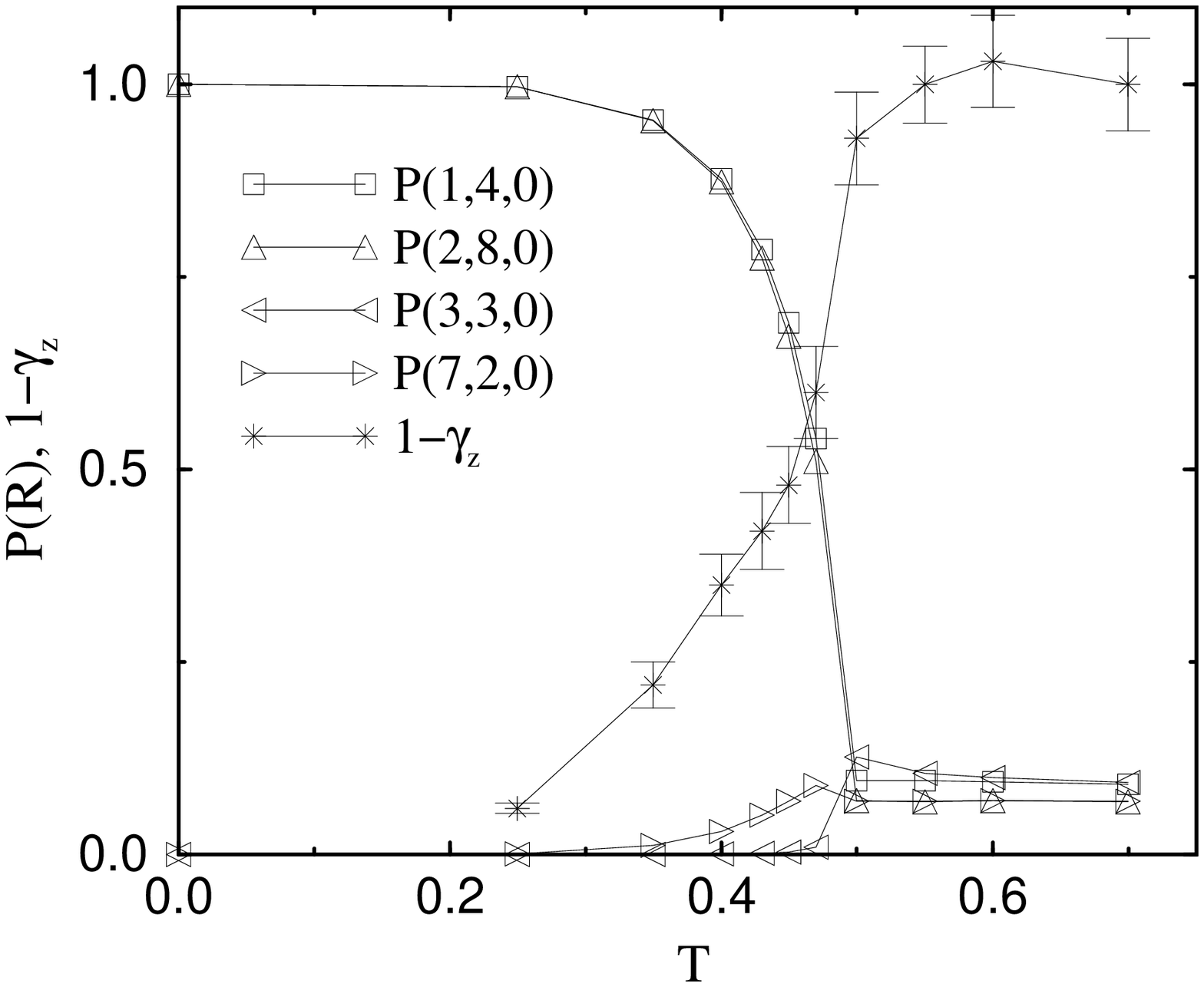,height=6.50cm,width=7.70cm}
{FIG.~7. \small Vortex-vortex spatial correlations $P(\bbox{R})$
for $R=0.04,$ $B=1/15,$ $\lambda_1=5,$ $L_x=L_y=30,$ $L_z=10$
as a function of temperature,
together with $1-\gamma_z$.}
\label{fig7}
\end{figure}
\begin{figure}
\centerline{
\psfig{figure=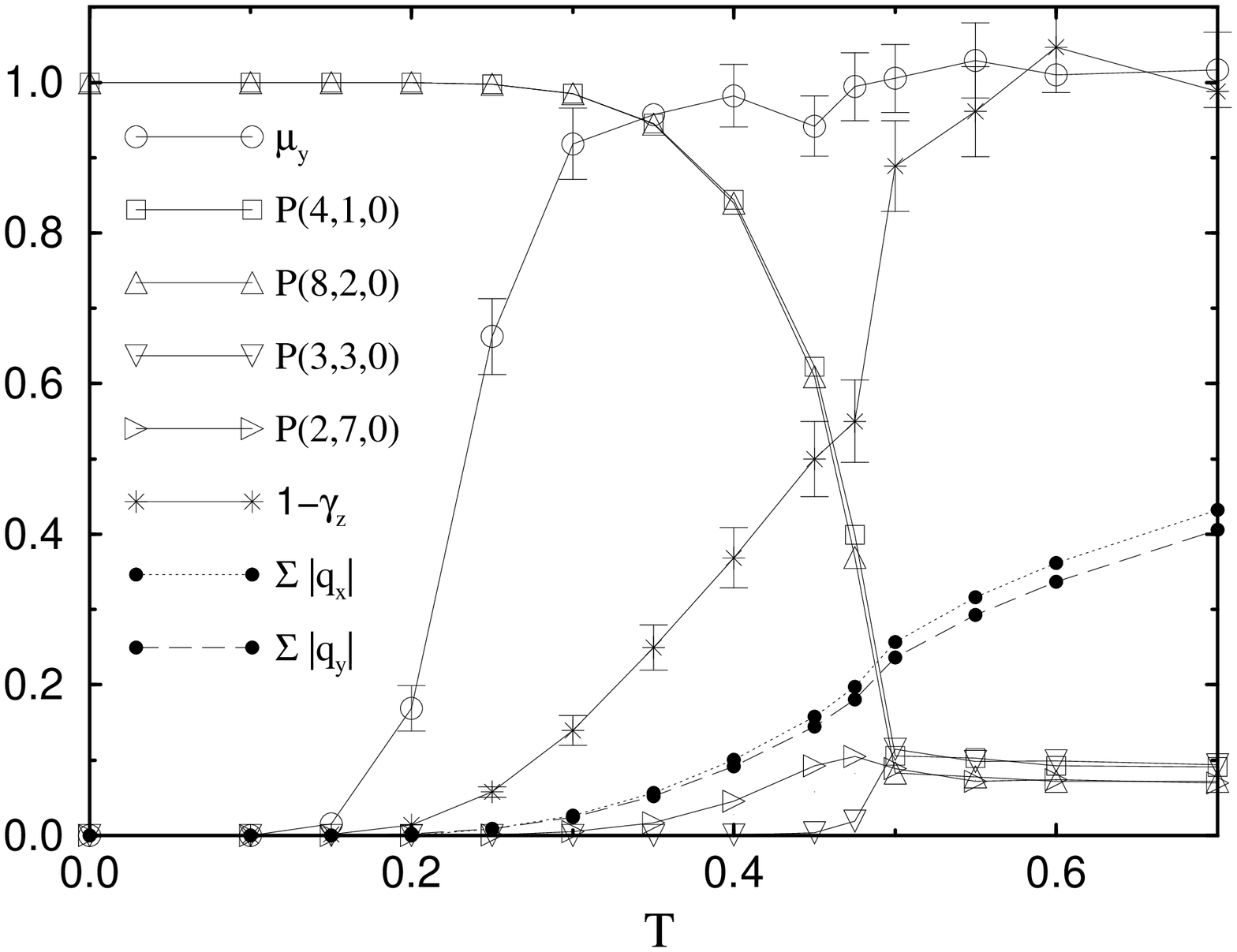,height=6.70cm,width=6.80cm}\hspace{0.3cm}
}
{FIG.~8. \small Vortex-vortex spatial correlations $P(\bbox{R})$
for $R=0.04$, $B=1/15$, $\lambda_1=5$, $L_x=L_y=L_z=15$ as a function of temperature,
together with $\mu_y$, $1-\gamma_z$, and the absolute vorticities
in $x$ and $y$ direction.
Sample error bars are shown for $\mu_y$ and $1-\gamma_z$.}
\label{fig8}
\end{figure}
\begin{figure}
\centerline{
\psfig{figure=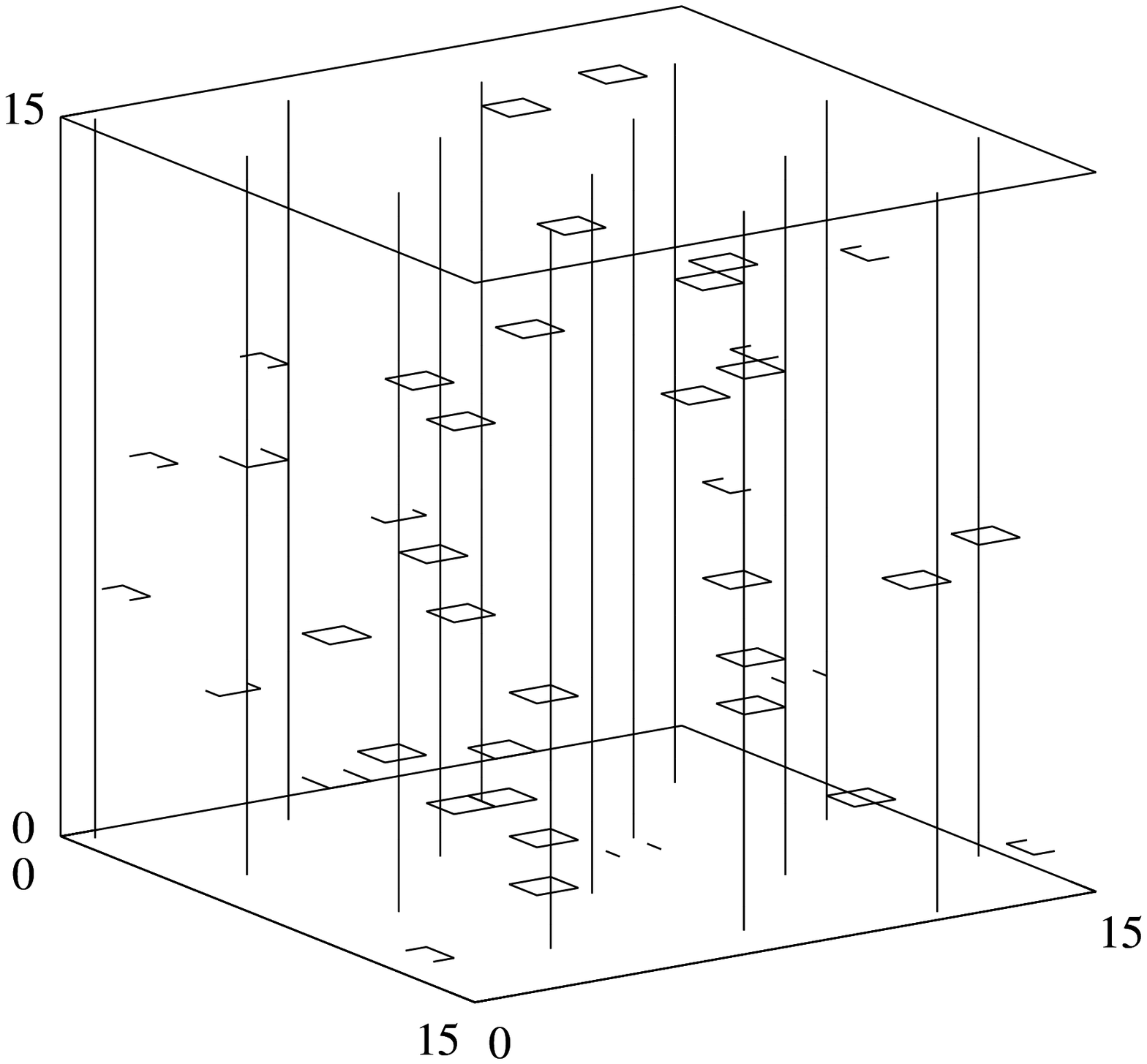,height=3.5cm,width=4.3cm}
}
{FIG.~9. \small Snapshot of the vortex-line configuration in the liquid phase
for $R=0.04,$ $B=1/15,$ $\lambda_1=5,$ $L_x=L_y=L_z=15,$ $T=0.25$.}
\label{fig9}
\end{figure}
Summarizing the above results, we find that $\mu_y$ generally tends to
rise at significantly lower temperatures than $1-\gamma_z$.
We will now present a consistency check on these response coefficients.
We do this by considering
the $B=0$ superconductor-to-normal transition that is driven by
(the unbinding of) thermally excited vortex loops.
In this transition, the loss of superconducting coherence
is accompanied by a peak in the specific heat and a sharp increase in the
number of vortex loops \cite{Dasgupta},
and the results for the response coefficients
should therefore match these features.

In Fig.~6 we present the
results for $\mu_y$ following Carneiro together with
the helicity modulus $\gamma_z$ following Chen and Teitel, calculated
in the same simulation for an isotropic system at $B=0$.
We find that $\mu_y$ rises at much lower temperatures than
$1-\gamma_z$. The peak in the
specific heat, an independent signature  of the transition from the
superconducting to the normal state, is situated around $T\approx5.9$ and
coincides with the rise
in $1-\gamma_z$. This provides strong evidence that $\gamma_z$ and not $\mu_y$
is the correct quantity to measure the superconducting coherence.
In Fig.~6 we also show the absolute vorticity per unit cell
as a function of temperature, which is a measure of
the number of thermally excited vortex loops.
The increase in the vorticity is strongest close to
$T \approx5.9$, which is another signature of the transition \cite{Kohring,CaPRB96}.
The rise in $\mu_y$ takes place at temperatures
where the number of vortex-loop fluctuations is still
very small and  where superconductivity can therefore
not yet be destroyed. As insets in Fig.~6 we show
snapshots of the vortex loop configurations for
$T=4$ and $T=5.7$.
We conclude that fluctuations in the net vorticity $q_y$
that are induced by the allowed incomplete loops
in the planes $x=1$ and $x=L_x$
and that lead to a nonzero $\mu_y$, do not signal a loss
of superconducting coherence in the bulk \cite{opm}.

\subsection{Anisotropic case.}
We now discuss results for a moderately
anisotropic case in a non-zero applied field.
In Figs.~7 and 8 we compare results for 
$R=1/25$ (or $\Gamma=5$), corresponding to
YBCO. Here we took $B=1/15$ and $\lambda_1=5$.
The curve with data points shown as stars in Fig.~7 is the
helicity modulus $\gamma_z$ following Chen and Teitel,
or more precisely $\gamma_z(k_x)$ at the lowest $k_x$.
The results for $1-\gamma_z(k_x)$ are shown in Fig.~11. 
The lattice size was taken to be $L_x=L_y=30, L_z=10$.
We observe  that for this anisotropic case, $1-\gamma_z$
rises to unity approximately
at the vortex-lattice melting 
temperature.
Similar results were obtained
in a recent work by Nguyen et al. \cite{Nguyen}, where
the effect of anisotropy was studied systematically.
Due to the anisotropy, the energy cost of vortex-loop fluctuations
parallel to the $x$-$y$ planes is reduced. 
As explained in Ref. \cite{Nguyen}, these parallel loops,
which are most easily nucleated
near a vortex line running in the $z$ direction,
contribute to the decoupling of the planes,
so that longitudinal coherence is lost at lower temperature
than for the isotropic case.
In Ref.~\cite{Nguyen} it was shown in addition that for strong anisotropy
$R\leq 1/100$, the parallel loops proliferate already at a
temperature (far) below the melting temperature of the vortex lattice,
thus giving rise to a
vortex lattice phase without longitudinal coherence
in between these temperatures.

\begin{figure}
\psfig{figure=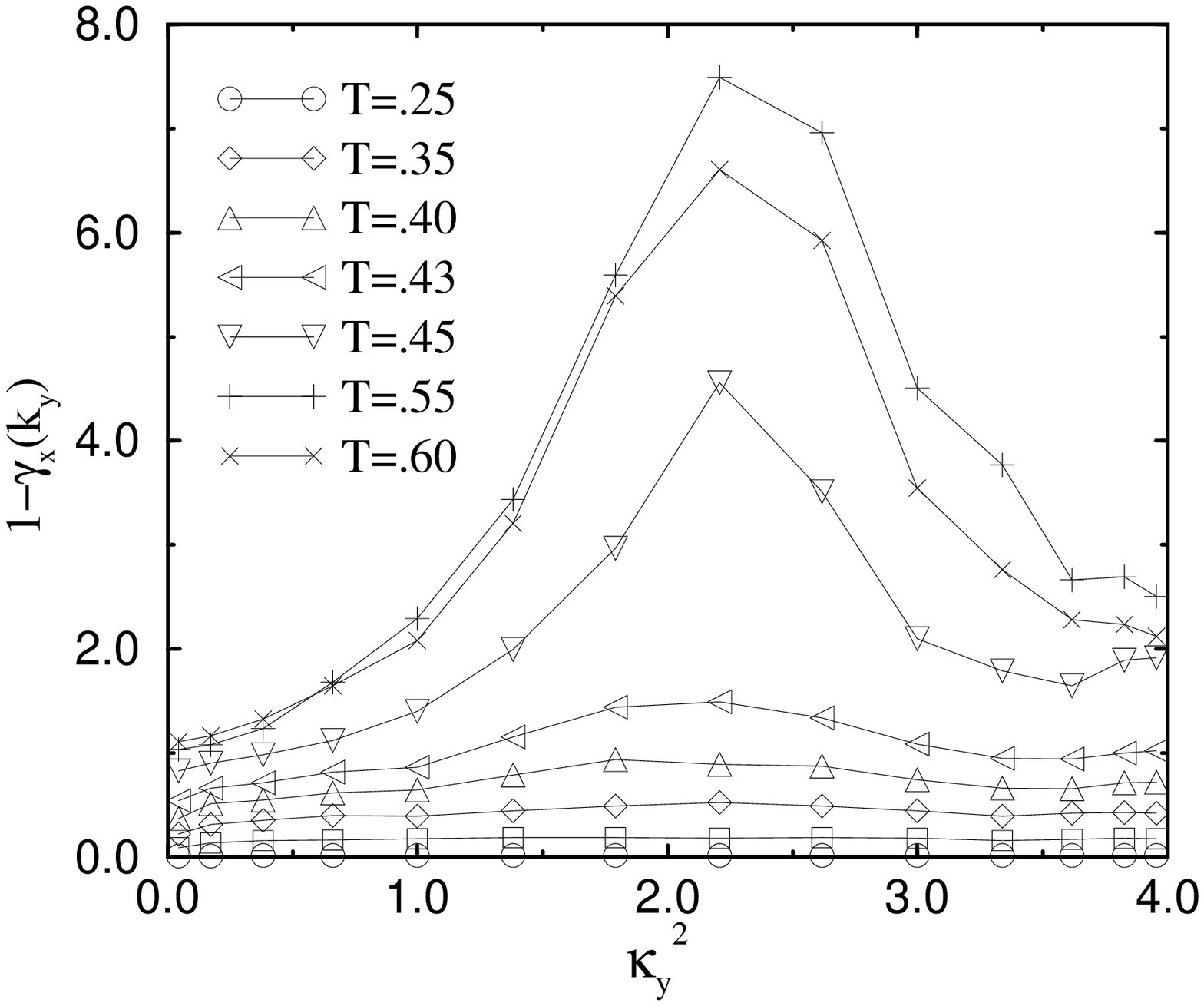,height=5.90cm,width=7.30cm}
{FIG.~10. \small Transverse response $1-\gamma_x(k_y)$ 
for different temperatures, calculated at constant $B$,
for $R=0.04,$ $B=1/15,$ $\lambda=5,$ $L_x=L_y=30,$ $L_z=10$.
The peak for higher T is at a wave vector commensurate with the vortex lattice.}
\label{fig10}
\end{figure}
\begin{figure}
\psfig{figure=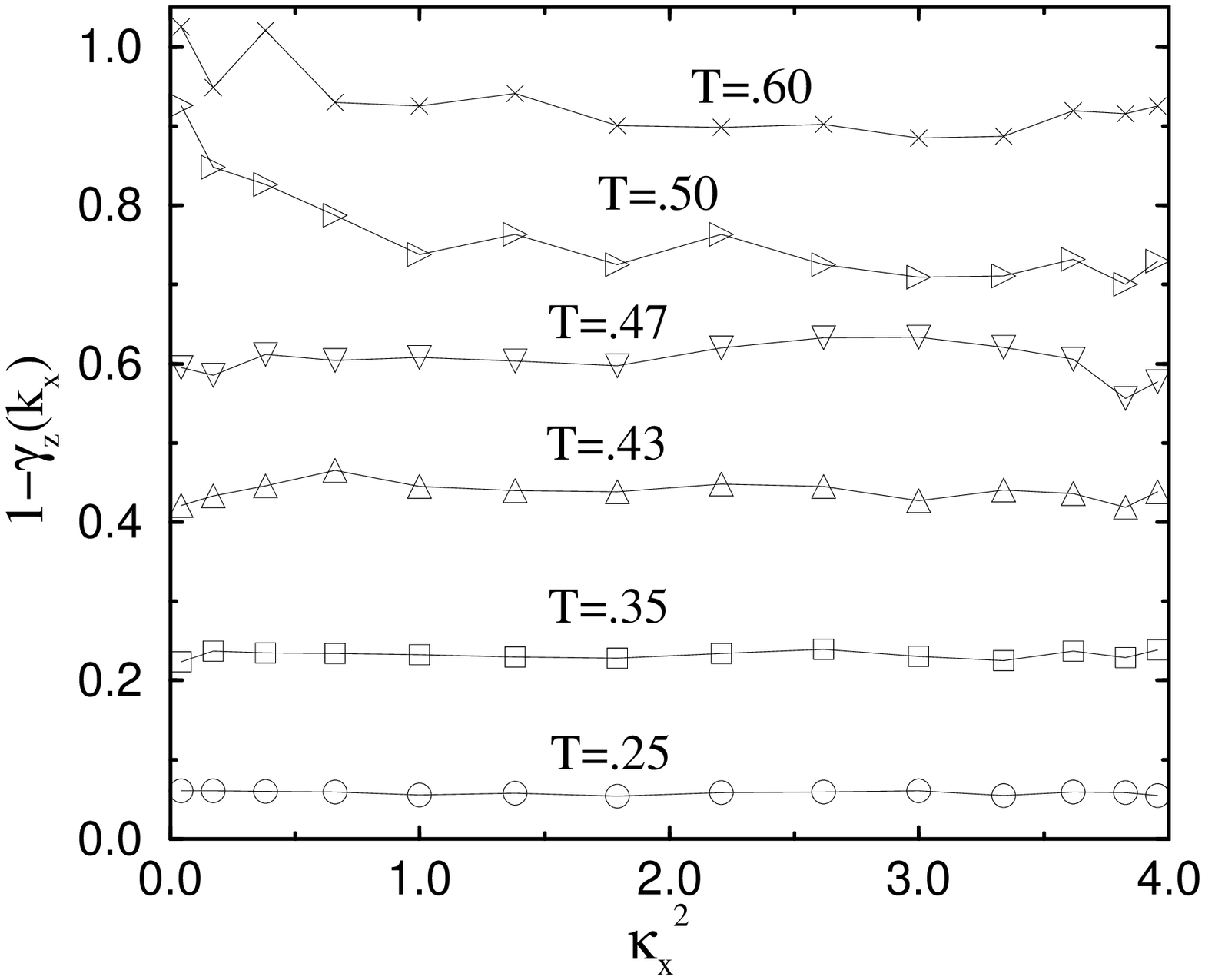,height=5.90cm,width=7.30cm}
{FIG.~11. \small Transverse response $1-\gamma_z(k_x)$ 
for different temperatures, calculated at constant $B$,
for $R=0.04,$ $B=1/15,$ $\lambda=5,$ $L_x=L_y=30,$ $L_z=10$.}
\label{fig11}
\end{figure}
In Fig.~8 we show Carneiro's permeability $\mu_y$ calculated
at the same parameters as considered in Fig.~7.
Here we used $L_x=L_y=L_z=15$. We observe that
there is a similar reduction of the temperature at which
longitudinal coherence is lost when compared to the
isotropic case. However, $\mu_y$ rises already well
below the melting temperature. 
The only physical mechanism  that could be
responsible for this early
rise would be  a proliferation
of parallel loops. However, there is no
such proliferation at this temperature, as can be seen from
the snapshot of the vorticity configuration at $T=0.25$
shown in Fig.~9. 
There are only few
parallel loops present, as can be seen 
from the absolute vorticities shown in Fig.~8
as well. 
However, according to the result for $\mu_y$, 
the planes should be almost decoupled at this temperature.
This again gives strong evidence that $\mu_y$ is  governed by artefacts.
In Fig.~8 we also show $1-\gamma_z$ calculated simultaneously
from the fluctuations of $q_x$ at finite $k_y$.  In agreement with
the calculations shown in Fig.~7, this longitudinal response
is found to turn resistive at the vortex-lattice melting temperature.

\section{Summary}

We investigated the occurrence of longitudinal superconductivity in the
vortex liquid phase in a clean type-II superconductor,
within the lattice London model, motivated by
recent contradicting results reported in Refs.
\cite{taochen} and \cite{CaPRL95}.
For an isotropic superconductor,
our results show
three  distinct regimes as a function of temperature.
Between the Abrikosov lattice phase
and a  vortex liquid without longitudinal
coherence,
there is a liquid regime
in which
longitudinal superconducting coherence extends over length scales
larger than the system thickness studied \cite{taochen}. 
In the context of flux-transformer experiments,
this intermediate regime should correspond to resistive regime with
$z$-correlated vortex motion for samples that are sufficiently thin.
In the moderately anisotropic
case $R=1/25$ (corresponding to YBCO), we find that
longitudinal superconducting
coherence is lost close to the flux-line lattice melting temperature.
This is due to a proliferation, close to the melting temperature,
of vortex loops between the layers \cite{Nguyen}.
In recent flux-transformer work \cite{Lopez} on 
YBCO single crystals, flux-line lattice melting and a loss 
of maximally $z$-correlated flux motion was found to coincide.

\section*{Acknowledgments}

We thank H.G. Evertz, M. Kraus, H. Nordborg and A. van Otterlo for discussions,
and G. Carneiro as well as A.K. Nguyen and A. Sudbo for sending us
their work prior to publication.
We gratefully acknowledge financial 
support by the ``Bayerischer Forschungsverbund Hochtemperatur-Supraleiter (FORSUPRA)".

\end{multicols}

\begin{references}
\bibitem{opmerking0} Thus the dominant $k_\nu$ dependence
of the vorticity correlation $\langle q_\sigma (k_\nu ) q_\sigma (-k_\nu )\rangle$
is described by the denominator in the second term in
Eq.~(\ref{gamma}). From the functional form
of this dependence one may indeed  infer that
an extrapolation to $k_\nu=0$ of the unnormalized
correlation  (i.e. the numerator in Eq.~(\ref{gamma})) using a polynomial fit
can only be successful when data points are available for
$k_\nu$ values significantly smaller than $1/(\lambda (1+\delta_{\mu ,z}(\Gamma -1)))$.
\bibitem{opmerking} In MC studies of a uniformly frustrated $XY$ model on a cubic lattice,
an intermediate disentangled vortex-liquid phase
was found for a magnetic frustration $f=1/25$ \cite{Yinghong}, but
not at $f=1/5$ \cite{Yinghong2} for this reason.
\bibitem{opmerking2} The decrease of $T_m$ with decreasing
$B$ is due to the weakening of the artificial pinning.
\bibitem{opm}We note that this
may be less apparent  from $B=0$ results for
smaller values of $\lambda$, such as considered in
$B=0$ calculations in Ref.~\cite{CaPRB96}.
\end{references}
\end{document}